\documentclass[a4paper,12pt]{article}
\usepackage{graphicx}

\begin{document}

\title{ Lattice Study of the High Density State of SU(2)-QCD }

\author  {  S. Muroya$^{1)}$,  A. Nakamura $^{2)}$ and  C. Nonaka$^{2)}$ \\
\vspace{.5cm} \\
     {\it $^{1)}$Tokuyama Women's Coll. Tokuyama, 745-8511, Japan}\\
     {\it $^{2)}$IMC, Hiroshima University,  Higashi-Hiroshima 739-8521, Japan }\\
}

\date{}
\maketitle

\begin{abstract}
We investigate high density state of SU(2) QCD by using Lattice QCD simulation 
with Wilson fermions.
The ratio of fermion determinants is evaluated at each step of the Metropolis
link update by Woodbury formula.   At $\beta=0.7$, and $\kappa = 0.150$, 
we calculate the baryon number density, the Polyakov lines, and the energy density of gluon sector with chemical potential $\mu$=0 to 0.8 on the $4^{3} \times 12$ lattice.   Behavior of the meson propagators and  diquark propagators with finite chemical potential are also investigated.

\vspace{1pc}
\end{abstract}

\section{INTRODUCTION}

High density state of the strongly interacting matter is attracting much attention \cite{Satz}
and one of the main targets of this workshop.  Numerical simulation  based on the 
 Lattice QCD 
is the established method to investigate the high temperature state of the strongly 
interacting matter; However, 
because of the well known problem that  chemical potential makes action 
 complex, the progress of the
lattice QCD in the {\it finite density} has been rather slow. 
Indeed, after the first dynamical quark simulation with the
chemical potential was done for SU(2)\cite{Nakamura84},
to our knowledge,  only few full SU(3) QCD calculations had been tried.
Recently, Fodor and Katz\cite{Fodor} proposed the nobel method to draw 
the critical line on the $T-\mu$ plane
and to find the point where first order phase transition turns to the crossover.   
Their method is based on the Lee-Yang Zero, therefore, it seems to work well only on 
the critical line and it seems still out of scope to investigate physics across 
the phase transition.

Quark in the real world is SU(3) fundamental representation; SU(2)-QCD and  
quark in the adjoint representation are simple toys for the theorist.  
However,   due to the recent progress in analytical investigations, 
we can hopefully obtain  some information on real SU(3)-QCD  
through the investigation of the finite density region of the "QCD-like" 
theories \cite{Kogut00}. 
The QCD-like theories, such as  SU(2)-QCD, quark model in the adjoint 
representation and QCD at finite isospin density, 
are expected to have less difficulties in numerical analyses.
In these years, there are indeed high activities in Monte Carlo 
calculations with dynamical quark of such kinds of 
models \cite{Lombardo99a}.
Furthermore, recent analyses on the color superconductivity suggest the possible
realization of SU(2) part of the color SU(3) as a residual interaction in the color superconducting state \cite{CCC}.
In this paper, we report our recent work on the SU(2)-QCD finite density states  with
Wilson fermions.

\section{Chemical Potential on the Lattice}

The chemical potential, $\mu$, is introduced in the fermion action,
$\bar{\psi} W \psi$, as \cite{Nakamura85},
\begin{eqnarray}
W(x,x') &=& \delta_{x,x'} - \kappa \sum_{i=1}^{3} \left\{ 
        (1-\gamma_i) U_i(x) \delta_{x',x+\hat{i} } 
      + (1+\gamma_i) U_i^{\dagger}(x') \delta_{x',x-\hat{i}} \right\} 
\nonumber \\
& &-  \kappa \left\{ 
        e^{+\mu a}(1-\gamma_4) U_4(x) \delta_{x',x+\hat{4}} 
      + e^{-\mu a}(1+\gamma_4) U_4^{\dagger}(x') \delta_{x',x-\hat{4}} \right\} .
\label{Wfermion}
\end{eqnarray} \noindent
Little is known about the behavior of dynamical fermion simulations
when the chemical potential is introduced.  
For $\mu \ne 0$, the relation 
$W^{\dagger}=\gamma_5 W \gamma_5$ 
does not hold, and hence $\mbox{det} W$ is in general not real. 

Since $U_\mu^{*} = \sigma_2 U_\mu \sigma_2$, the fermion matrix for SU(2)
has the following propertiy:
\begin{eqnarray}
W(x,x'; \gamma_\mu)^{*} &=& \sigma_2 ( 
         \delta_{x,x'} - \kappa \sum_{i=1}^{3} \left\{ 
        (1-\gamma_i^{*}) U_i(x) \delta_{x',x+\hat{i} } 
      + (1+\gamma_i^{*}) U_i^{\dagger}(x') \delta_{x',x-\hat{i}} \right\}
\nonumber \\
& &-  \kappa \left\{ 
        e^{+\mu a}(1-\gamma_4^{*}) U_4(x) \delta_{x',x+\hat{4}} 
      + e^{-\mu a}(1+\gamma_4^{*}) U_4^{\dagger}(x') \delta_{x',x-\hat{4}} \right\}
         ) \sigma_{2},
\nonumber \\
&=& \sigma_2 W(x,x'; \gamma_\mu^{*}) \sigma_2 .
\label{Wfermion-SU2}
\end{eqnarray} \noindent
Then $\{ \mbox{det} W(x,x'; \gamma_\mu) \}^{*} 
= \mbox{det}  W(x,x'; \gamma_\mu^{*})$.
$\gamma_\mu^{*}$  belong a representation which also satisfies the anti-commutation 
relations same as $\gamma_\mu$ 
and $\mbox{det} W$ should not depend on the representation of $\gamma$-matrix.
Therefore, differing essentially from the SU(3) case, the action of the SU(2)-QCD is real 
with chemical potential. 

However,  numerical simulation is not straightforward and instability occurs 
with large chemical potential, which 
makes lattice simulation difficult \cite{Muroya01}.
Therefore, we need careful treatment for the updation 
of the configulation.
We here adopt locally updating exact algorithm
based on the Woodbery formula \cite{Nakamura84}. 
The algorithm is summarized in the appendix.


\section{Thermodynamical Quantities}

First, we calculate thermodynamical quantities, such as Polyakov line, gluon energy density and baryon number density with $4^{3} \times 12$ lattice.  We used Wilson fermion and Iwasaki improved action.

The expectation value of the baryon number density is given by,
\begin{equation}
<n> = \frac{1}{\beta V_s} \frac{\partial}{\partial\mu} \log Z
\end{equation} \noindent
where $V_s$ is the spatial volume $N_{x} \times N_{y} \times N_{z}$.
Energy density, $\varepsilon$, is given as, 
\begin{equation}
\varepsilon = \frac{1}{V_s}\left(- \frac{\partial}{\partial \beta} 
+ \frac{\mu}{\beta} \frac{\partial}{\partial \mu} \right) \log Z
\end{equation}
The derivative of the partition function is composed of two parts,
\begin{equation}
(\log Z)'= \frac{1}{Z} \int {\cal D}U{\cal D}{\bar \psi}{\cal D}{\psi}
(-S'_{G}-S'_{F})e^{(-S_{G}-S_{F})},
\end{equation}
where $S_{G}$ and $S_{F}$ are gluon action and fermion action, respectively.
We denote the contribution of the gluon action part (first term in the r.h.s. of eq.(3)) by  gluon energy density.
Figure 1 displays  Polyakov line, gluon energy density and baryon number density 
of $\mu$ where $\beta = 0.7$ and $\kappa = 0.15$, respectively.  
All quantities start to have non-zero value at about $\mu=0.4$ and rise up with
chemical potential in the region $0.4< \mu < 0.8$.
None of them is order parameter of the phase transition in the exact sense, and
since the lattice size is small, no sharp change is seen. 
However,  growing up  of the these quantities  indicates that quarks and gluons
become free from the
confinement force at finite chemical potential about $0.4< \mu < 0.8$.

%
\begin{figure}
\begin{center}
\includegraphics[width=.9 \linewidth]{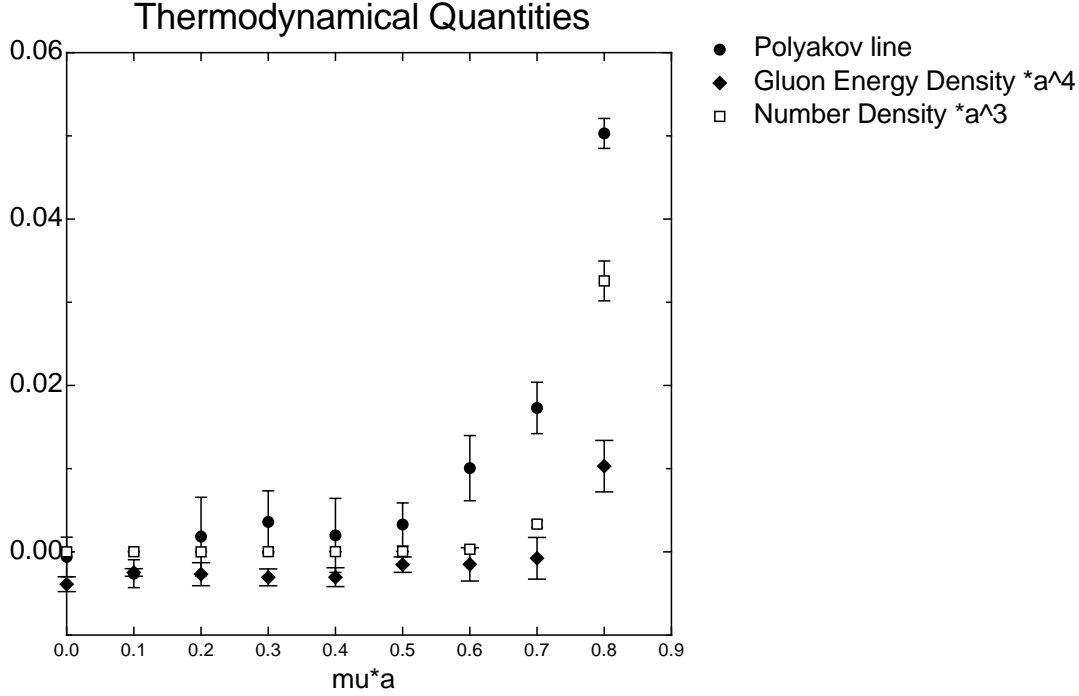}
\end{center}
\caption{Polyakov line, gluon energy density, and baryon number density as a function of  chemical Potential $\mu$.}
\label{Fig-density}
\end{figure}


\section{Meson and Baryon}


Because fundamental representation 
of the SU(2) is {\bf 2}, baryon in the SU(2)-QCD is diquark state.  Scalar diquark state 
and pseudo-scalar diquark state are given as $(C^{-1}\psi)^{T}\gamma_{5} \psi$ and 
$(C^{-1}\psi)^{T} {\bf1} \psi $, respectively, with $C$ being charge conjugation 
matrix.  Charge  conjugation makes transform property of the diquark state look
opposite  to the ordinary  combination   of the 
 scalar  $\bar{\psi} {\bf1} \psi$  and pseudo scalar $ \bar{\psi} \gamma_{5} \psi$.  
We denote  scalar diquark state 
and pseudo-scalar diquark state as b5b and b1b, respectively, for the abbreviation. 

\begin{center}
\begin{figure}[thb]

\begin{center}

\begin{minipage}{ 0.42\linewidth}
\includegraphics[width= \linewidth]{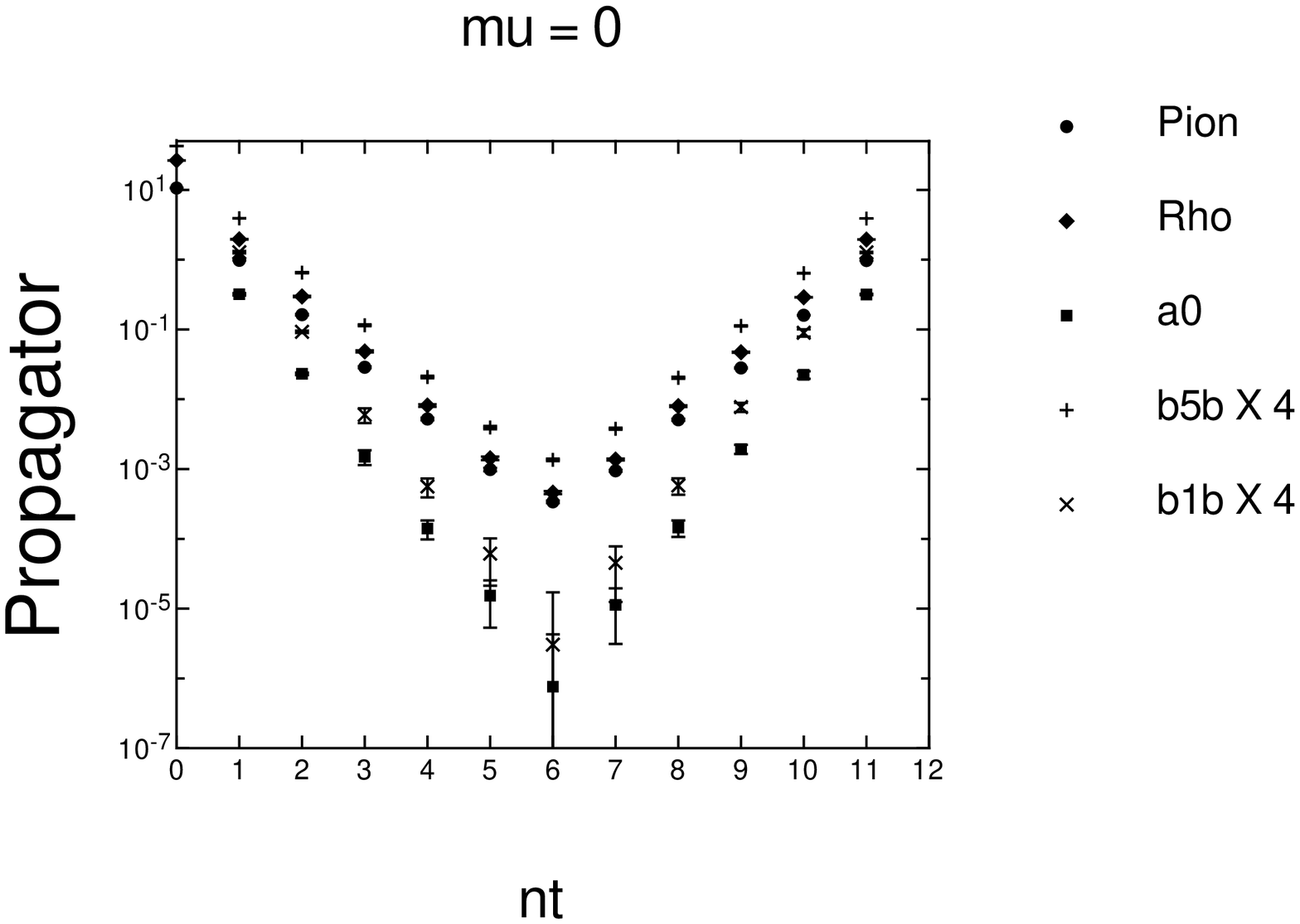}
\end{minipage}
\begin{minipage}{ 0.42\linewidth}
\includegraphics[width= \linewidth]{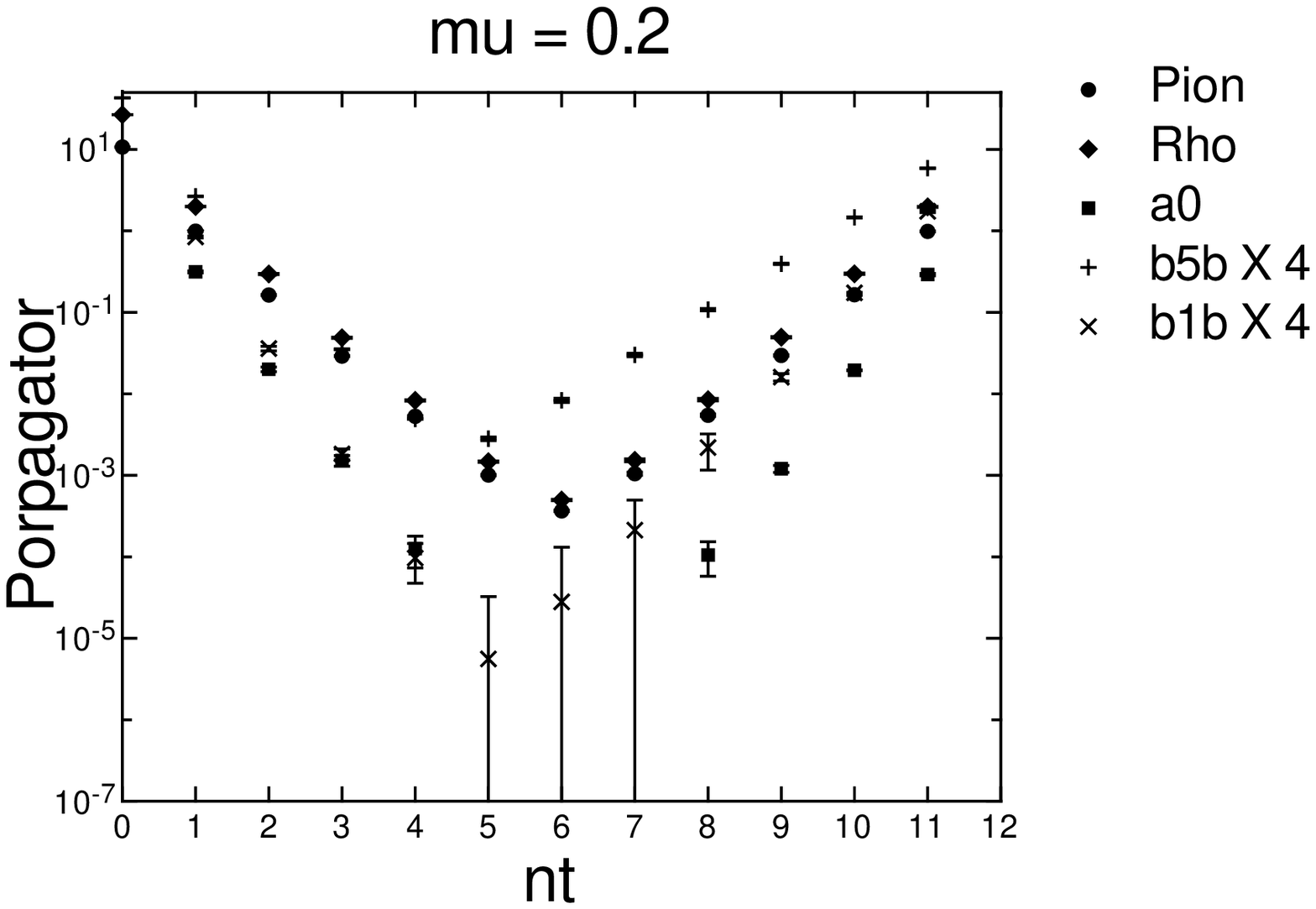}
\end{minipage}
\end{center}
\begin{center}

\begin{minipage}{ 0.42\linewidth}
\includegraphics[width= \linewidth]{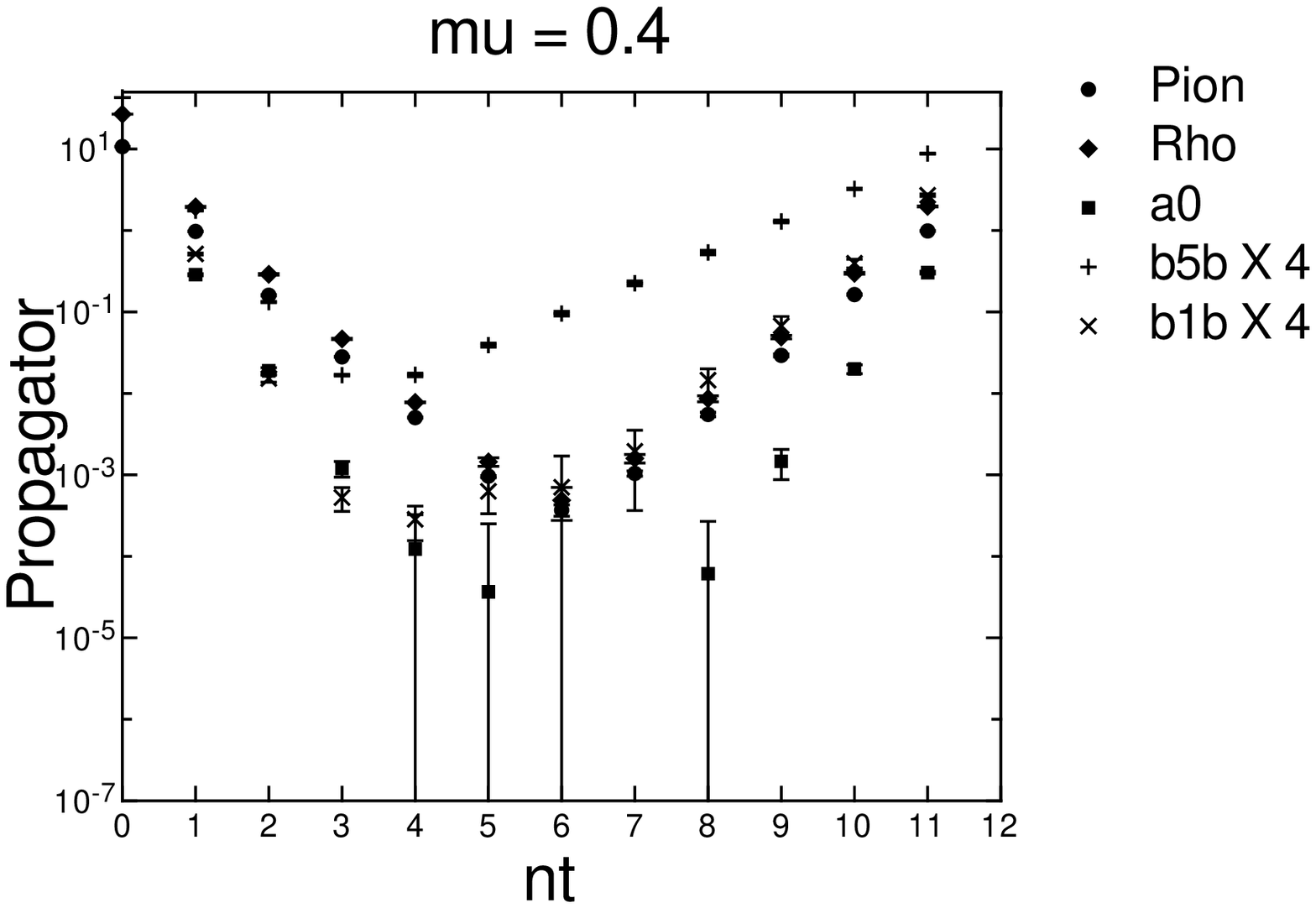}
\end{minipage}
\begin{minipage}{ 0.42\linewidth}
\includegraphics[width= \linewidth]{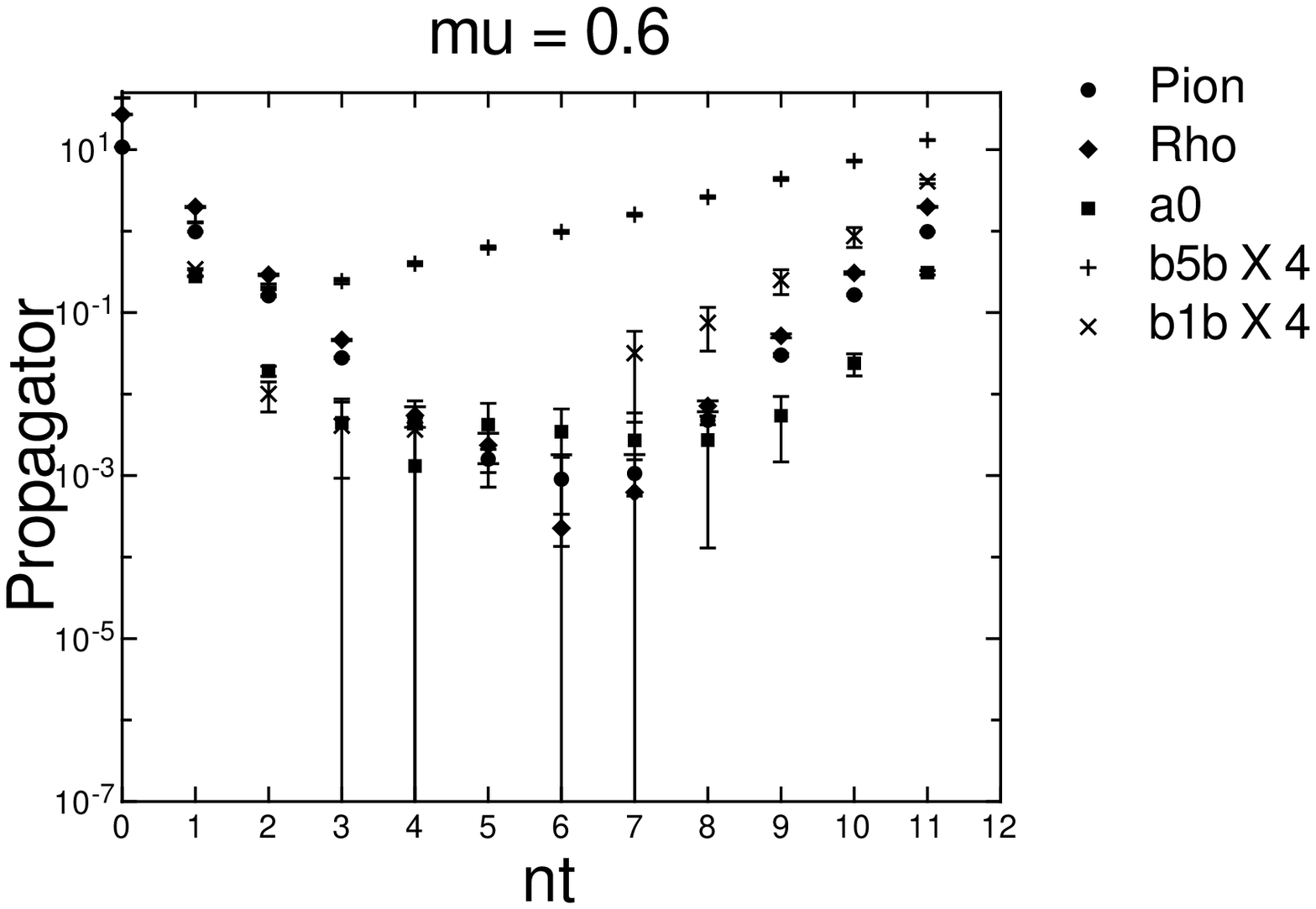}
\end{minipage}

\end{center}
\begin{center}

\begin{minipage}{ 0.42\linewidth}
\includegraphics[width= \linewidth]{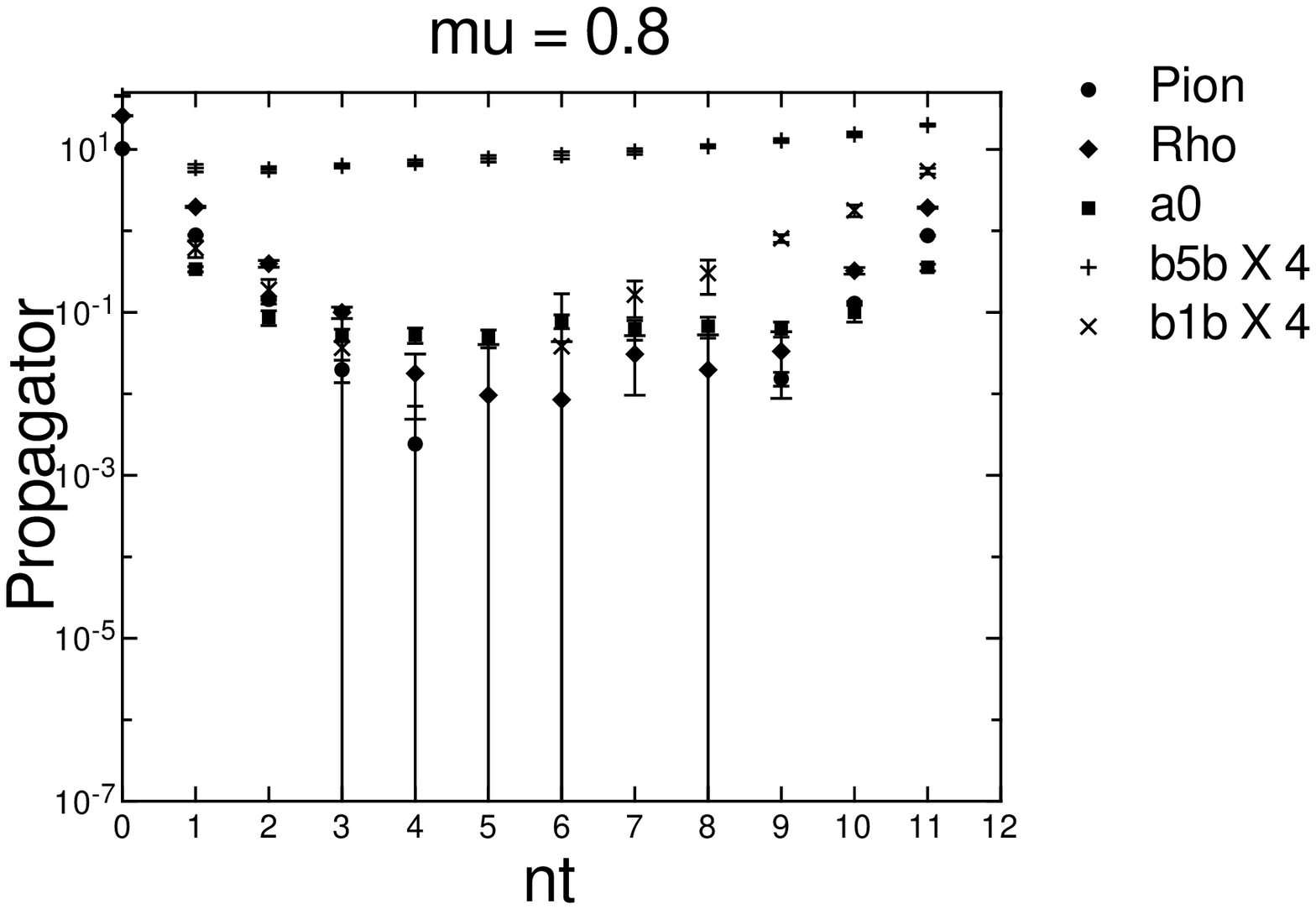}
\end{minipage}
\begin{minipage}{ 0.42\linewidth} 
\end{minipage}

\end{center}

\caption{Propagators of mesons and diquarks of $4^{3}\times 12$ lattice at $\beta = 0.70$ and $\kappa = 0.150$. In order to avoid overplot of the data,  diquark propagators are shifted 4 times. }
\label{Fig_propa}
\end{figure}
\end{center}

We evaluate propagators of
the pseudo scalar iso vector meson,  $ \pi$, the scalar iso vector meson,  $ a_0$, 
vector meson, $\rho$, pseudo scalar diquark (b1b) and scalar diquark (b5b).  With 
 vanishing chemical potential,  the correlator of the $\pi$ and scalar diquark 
 degenerate and  so do $a_0$ and  pseudo scalar diquark.  
 Because meson has no net baryon number, effect of the chemical potential is expected to appear in the meson propagator only through the mass. On the other hands, baryon (diquark) has definite baryon number; Hence, in addition to the change of the mass, the affect of the finite chemical potential on the particle and antiparticle is  
in opposite sign, which causes the asymmetry of the propagator in time direction.  
 With a  finite chemical potential $\mu$, propagator of diquark, 
$G_{b1b}(x,x')$ and $G_{b5b}(x,x')$,  should behave as, 
$$
G_{b*b}(t,\mu) = G_{b*b}(T-t,-\mu) ,
$$
in  contrast to the meson propagator,
$$
G_{m}(t,\mu) = G_{m}(T-t,\mu)
$$
with $T$ being lattice size in the time direction.

Figure \ref{Fig_propa}  shows propagators of mesons and diquarks.  At $\mu = 0$,  as expected, pseudo scalar $\pi$ and scalar diquark (b5b) and scalar $a_0$ and  pseudo scalar diquark coincide, respectively (in Fig.\ref{Fig_propa}, to avoid the overplot, diquark propagators are shifted by factor 4).  With finite chemical potential, asymmetry of the diquark propagators in time direction and anti-time direction become stronger with $\mu$.  On the other hand, meson propagators keeps symmetric in $n_t$.  Hence, in our calculation, effect of the finite chemical potential works appropriately
on the hadron propagators.  Hands \cite{Hands99}, reported that with finite 
chemical potential, 
 propagator of the  scalar diquark (b5b) and pseudo scalar diquark (b1b) become parallel 
and they proposed the interpretation that  both diquark states give the same spectra 
and the difference corresponds to the diquark condensation.  However, at least the 
 present statistics,  our results do not give the the diquark propagators in 
parallel.

\section{CONCLUDING REMARKS}

We present numerical study of SU(2)-QCD with the
chemical potential on lattice with Wilson fermions.
Although the lattice is not large, behaviors of the thermodynamical 
quantities suggest that 
we are at around the confinement/deconfinement phase
transition.
Though the change of the propagator as a function of the chemical potential 
$\mu$ is almost consistent with the results of \cite{Hands99}, however,  
the spectrum of the diquarks seem to be different.   
But we need more statistics to conclude definite 
results.  Estimation of the mass as a function of the chemical potential in the chiral
limit is now  in progress.

In our calculation, numerical convergence becomes worse and worse with larger 
chemical potential and at $\mu = 1.0 $ vanishing determinant makes simulation 
break-down.  We are adopting  the algorithm based on the exact calculation of the
ratio of fermion determinant.  Therefore,  we can analyze the change of the 
distribution of eigenvalues of the fermion matrix  with finite chemical potential \cite{Muroya01}.   Investigation of the physical meaning of the numerical 
 instability and distribution of the  eigenvalue are also our next task.

\section{Acknowledgment}

This work is supported by Grant-in-Aide for Scientific Research by
Monbu-Kagaku-sho, Japan (No.11440080 and No. 12554008).
Simulations were performed on SR8000 at IMC, Hiroshima
University, SX5 at RCNP, Osaka university, SR8000 at KEK and
VPP5000 at Science Information Processing Center, Tsukuba university.

\section{Appendix}
We  adopt an algorithm where the ratio of the determinant,
\begin{equation}
 \frac{ \mbox{det} W(U+\Delta U) }{ \mbox{det} W(U) } 
= \mbox{det} (I + W(U)^{-1} \Delta W)
\end{equation} \noindent
is evaluated explicitly at each Metropolis update process, 
$U \rightarrow U+\Delta U$, 
where 
$\Delta W \equiv W(U+\Delta U) - W(U)$ \cite{Barbour87}.  
An essential ingredient of the algorithm is Woodbury formula, 
\begin{equation}
(W+\Delta W)^{-1} 
= W^{-1}- W^{-1}\Delta W
( I + W^{-1}\Delta W  )^{-1} W^{-1}.  
\label{Woodbury}
\end{equation}

Suppose we update link variables $U_\mu(x)$s only on
a subset $H$ of whole lattice.  Though $\Delta W \neq 0$
only on $H$,  Woodbury formula (\ref{Woodbury}) still holds
on $H$, and  in this case, we can get
the ratio of the fermion determinant as far as $U_\mu(x)$s
are locally updated only inside $H$.  We take a $2^4$ hypercube as $H$.
When we move to the next hypercube, $(W^{-1})_H$'s
are initialized by CG method.

We employ an algorithm which takes into account
the ratio of fermion determinant exactly, and has large
Markov step, but we suffer from numerical instability at about $\mu=1.0$

\begin{small}

\end{small}


\begin{thebibliography}{99}
\bibitem{Satz} H.~Satz, hep-ph/0009099. 
\bibitem{Nakamura84} A.~Nakamura, Phys.\ Lett., 149B (1984) 391.
\bibitem{Fodor}Z.\ Fodor and S.\ D.\ Katz,hep-lat/0106002. 
\bibitem{Kogut00} J.~B.~Kogut et al., Nucl.\ Phys.\ B582 (2000) 477;
D.~T.~Son and M.~A.~Stephanov, hep-ph/0005225,
J.~B.~Kogut, M.~A.~Stephanov and D.~Toublan, Phys.\ Lett.\ B464 (1999) 183.
\bibitem{Lombardo99a} M.-P. Lombardo, hep-lat/9907025;
           hep-lat/9906006.
\bibitem{Hands99} 
S.~Hands, J.~B.~Kogut, M-P.~Lombardo and S.~E.~Morrison,
                     Nucl.\ Phys.\ B558 (1999) 327; 
S.~Morrison and S.~Hands, hep-lat/9902012;
S.~Hands et al.,  hep-lat/0006018;
S.~J.~Hands, J.~B.~Kogut, S.~E.~Morrison, D.~K.~Sinclair,
                   hep-lat/0010028. 
%
\bibitem{CCC}K.\ Rajagopal and F. Wilczek, hep-ph/0011333.
\bibitem{Nakamura85} A.~Nakamura, Acta.\  Phys.\  Pol.\  B16 (1985) 635; 
P.~Hasenfratz and F.~Karsch, Phys.\ Lett, 125B (1983) 308. 
\bibitem{Muroya01} S.~Muroya, A.~Nakamura and C.~Nonaka, Nucl.\ Phys., 
Nucl.\ Phys.\ B Proc.\ Suppl.\ 94(2001)469.
\bibitem{Barbour87} I.~Barbour et al., J.\ Comput.\ Phys.\ 68 (1987) 227;  
A.~Nakamura et al., Comm.\ Phys.\ Comm.\ 51 (1988) 301; 
Ph.~de Forcrand et al., Phys.\ Rev.\ Lett., 58 (1987) 2011.
\end{thebibliography}
\end{document}